\documentclass[]{aastex631}

	\usepackage{hyperref}
   \usepackage{longtable}
\begin{document}

  \title{Comprehensive Parameter Determination of Exoplanet through \\
            Asteroseismic Host Star Constraints}

  \correspondingauthor{Sheng-Bang Qian}
		\email{qiansb@ynu.edu.cn}

		\author[0000-0002-8276-5634]{Wen-Xu Lin}
		\affiliation{Yunnan Observatories, Chinese Academy of Sciences, Kunming 650216, People's Republic of China}
		\affiliation{Department of Astronomy, School of Physics and Astronomy, Yunnan University, Kunming 650091, P. R. China}
		\affiliation{Key Laboratory of Astroparticle Physics of Yunnan Province, Yunnan University, Kunming 650091, P. R. China}
		\affiliation{University of Chinese Academy of Sciences, No.1 Yanqihu East Rd, Huairou District, Beijing, 101408, People's Republic of China}

		\author[0000-0002-5995-0794]{Sheng-Bang Qian}
		\affiliation{Department of Astronomy, School of Physics and Astronomy, Yunnan University, Kunming 650091, P. R. China}
		\affiliation{Key Laboratory of Astroparticle Physics of Yunnan Province, Yunnan University, Kunming 650091, P. R. China}

		\author[0000-0002-0796-7009]{Li-Ying Zhu}
		\affiliation{Yunnan Observatories, Chinese Academy of Sciences, Kunming 650216, People's Republic of China}
		\affiliation{University of Chinese Academy of Sciences, No.1 Yanqihu East Rd, Huairou District, Beijing, 101408, People's Republic of China}

      \author[0000-0001-9346-9876]{Wen-Ping Liao}
      \affiliation{Yunnan Observatories, Chinese Academy of Sciences, Kunming 650216, People's Republic of China}
      \affiliation{University of Chinese Academy of Sciences, No.1 Yanqihu East Rd, Huairou District, Beijing, 101408, People's Republic of China}

      \author[0000-0002-0285-6051]{Fu-Xing Li}
      \affiliation{Department of Astronomy, School of Physics and Astronomy, Yunnan University, Kunming 650091, P. R. China}
      \affiliation{Key Laboratory of Astroparticle Physics of Yunnan Province, Yunnan University, Kunming 650091, P. R. China}

      \author[0000-0002-5038-5952]{Xiang-Dong Shi}
      \affiliation{Yunnan Observatories, Chinese Academy of Sciences, Kunming 650216, People's Republic of China}
      \affiliation{University of Chinese Academy of Sciences, No.1 Yanqihu East Rd, Huairou District, Beijing, 101408, People's Republic of China}

      \author[0000-0003-3767-6939]{Lin-Jia Li}
      \affiliation{Yunnan Observatories, Chinese Academy of Sciences, Kunming 650216, People's Republic of China}
		\affiliation{University of Chinese Academy of Sciences, No.1 Yanqihu East Rd, Huairou District, Beijing, 101408, People's Republic of China}

      \author{Er-Gang Zhao}
      \affiliation{Yunnan Observatories, Chinese Academy of Sciences, Kunming 650216, People's Republic of China}
		\affiliation{University of Chinese Academy of Sciences, No.1 Yanqihu East Rd, Huairou District, Beijing, 101408, People's Republic of China}

  \begin{abstract}

   This study develops a robust framework for exoplanet characterization by leveraging asteroseismic constraints on host stars. Using precise photometric data from missions such as \textit{Kepler} and \textit{TESS}, we derive stellar parameters, including mass, radius, and age, with high accuracy through asteroseismic analysis. These stellar parameters are incorporated as priors in a Bayesian framework to refine planetary properties such as mass, radius, and orbital parameters. By applying Markov Chain Monte Carlo (MCMC) methods, we extract posterior distributions of planetary parameters, achieving significant improvements in precision and reliability. This approach is particularly effective for systems with evolved host stars, where precise stellar properties are essential for resolving uncertainties in planetary characterization. The results demonstrate the importance of asteroseismology in bridging stellar astrophysics and exoplanet science, enabling detailed studies of planetary system architectures and their dependence on host star properties. Our methodology underscores the synergy between stellar and planetary studies, paving the way for future research on exoplanet populations. This work provides a foundation for utilizing data from upcoming missions like \textit{PLATO}, ensuring continued advancements in the precision and scope of exoplanet characterization.

  \end{abstract}


  \section{Introduction} \label{sec:intro}

   The discovery and characterization of exoplanets have undergone a dramatic transformation with the advent of space-based telescopes. Missions such as \textit{Kepler} \citep{borucki2010kepler} and Transiting exoplanet survey satellite (\textit{TESS} \citet{ricker2015transiting}) have revolutionized our understanding of planetary systems, enabling the detection of thousands of exoplanets and providing precise light curves for both transit and asteroseismic analyses. The continuous advancements in photometric precision have not only facilitated the detection of Earth-like planets but also enabled detailed studies of their host stars, an essential component for accurate planetary characterization \citep{howard2012planet, fulton2018california}. With the forthcoming \textit{PLATO} mission \citep{rauer2014plato}, the synergy between exoplanet and asteroseismic research is expected to reach unprecedented levels, offering new insights into the formation, structure, and evolution of planetary systems.

   Asteroseismology, the study of stellar oscillations, provides a unique tool for probing the internal structures of stars \citep{chaplin2013asteroseismology}. By analyzing oscillation frequencies, it is possible to derive fundamental stellar parameters such as mass, radius, and age with exceptional precision \citep{aerts2010asteroseismology, metcalfe2010precise}. The advent of space telescopes capable of long-term, high-precision photometric monitoring has made asteroseismology a cornerstone of stellar astrophysics. For example, \textit{Kepler} data revealed oscillations in thousands of solar-like stars, leading to breakthroughs in stellar evolution studies and the calibration of scaling relations for red giants and subgiants \citep{stello2022extension}. Similarly, \textit{TESS} continues to expand this database, providing shorter-duration observations but across a wider field of view \citep{campante2016asteroseismic}.

   In parallel, the field of exoplanet science has flourished, with transit photometry and radial velocity measurements serving as the primary detection methods \citep{mayor1995jupiter, charbonneau1999detection}. The physical and orbital properties of exoplanets, such as radius, mass, and semi-major axis, depend critically on the properties of their host stars. Inaccurate stellar parameters can propagate into large uncertainties in derived planetary parameters, underscoring the need for precise stellar characterization \citep{seager2003unique, seager2007mass}. Asteroseismology provides a solution to this challenge by offering robust stellar parameters, which serve as essential priors for modeling planetary systems. This synergy has led to significant progress in determining the population statistics of exoplanets and understanding their physical diversity \citep{seager2010exoplanet,petigura2013prevalence}.

   The integration of asteroseismology into exoplanet studies represents a critical advancement in the field. By combining transit data with asteroseismic analysis, it is possible to achieve a high level of precision in both stellar and planetary characterization \citep{huber2013fundamental, silva2015ages, 2024AJ....168...27L}. For instance, studies of planets orbiting subgiant and red giant stars have revealed insights into their dynamical evolution and potential engulfment during stellar expansion \citep{matsumura2010tidal, mordasini2012characterization,hon2023close,lin2024revealing}. Moreover, asteroseismology plays a pivotal role in identifying the evolutionary stage of host stars, which is crucial for understanding the long-term stability and habitability of orbiting planets \citep{kasting1997habitable, pepe2011harps}. As observational capabilities improve, the comprehensive analysis of host stars and their planets will shed light on the physical processes governing planet-star interactions and the ultimate fate of planetary systems.

   This study focuses on combining asteroseismic constraints with transit and spectroscopic data to determine the physical properties of host stars and their exoplanets. By leveraging the extensive photometric datasets from \textit{Kepler} and \textit{TESS}, we derive precise asteroseismic parameters, such as the large frequency separation ($\Delta \nu$) and the frequency of maximum power ($\nu_{\text{max}}$). These parameters are then used to calculate stellar masses and radii, which serve as inputs for determining planetary properties. This integrated approach not only enhances the precision of planetary characterization but also demonstrates the transformative potential of asteroseismology in advancing exoplanetary science. The methods and results presented here highlight the growing intersection of these two fields and pave the way for future research in the era of next-generation space telescopes.

  \section{Calculation of Global Asteroseismic Parameters} \label{sec:GAP}

  We utilized the Python package $lightkurve$ \citep{2018ascl.soft12013L} to retrieve all accessible short-cadence and long-cadence light curves of Kepler-278 and Kepler-643 observed by \textit{Kepler} and 2-minute cadence light curves of HD 167042 observed by \textit{TESS} from the Mikulski Archive for Space Telescopes (MAST)\footnote{https://mast.stsci.edu/}. The Lomb-Scargle method \citep{lomb1976least, scargle1982studies} was then employed to analyze these light curves and extract the corresponding power spectra.
  It is important to note that for light curves containing transit signals (e.g., Kepler-278 and Kepler-643), their power spectra exhibit significant signals corresponding to the orbital period and its harmonics. However, these frequencies are much lower than those of the solar-like oscillation signals of the host stars, and thus they do not affect the analysis of the solar-like oscillations.

  We employed the \texttt{pySYD} package \citep{2022JOSS....7.3331C} to estimate the large frequency separation, $\Delta\nu$, and the frequency of maximum power, $\nu_{max}$. To create the signal-to-noise ratio (SNR) diagram, the estimated background noise was subtracted from the power spectrum, with the background noise determined using a smoothing filter of width $log_{10}(0.01 \mu Hz)$. The resulting SNR data were analyzed with the \texttt{PBjam}\footnote[2]{https://github.com/grd349/PBjam} package \citep{2021AJ....161...62N}, allowing us to identify the frequencies of the observed radial ($\nu_{n,0}$) and quadrupolar ($\nu_{n,2}$) oscillation modes of the host stars (refer to Table \ref{tab:nu0,nu2} in the appendix). For robust results, it is essential to adopt appropriate prior distributions for global asteroseismic parameters and effective temperature in \texttt{PBjam}, as these priors help to constrain the parameter space effectively. 

  To derive the effective temperature and other parameters of the host stars, we utilized the results from the Large Sky Area Multi-object Fiber Spectroscopic Telescope \citep[LAMOST;][]{wang1996special,su2004active,2012RAA....12.1197C,zhao2012lamost}, based on its Data Release 11 \footnote[3]{https://www.lamost.org/dr11/}. The atmospheric parameters for each star are detailed in Table \ref{tab:para_all}. However, since HD 167042 lacks LAMOST spectral data, we instead referenced the effective temperature and its associated uncertainty from \citet{2021ApJS..255....8R}. Accordingly, we combined the global asteroseismic parameters obtained from \texttt{pySYD} with the effective temperatures and their reasonable error margins to construct prior distributions. It is crucial to highlight that the global asteroseismic parameters obtained from \texttt{pySYD} differ from the posterior distributions generated by \texttt{PBjam} (e.g., the posterior median value of $\Delta\nu$ displayed on the x-axis of Figures \ref{fig:echelle278} \ref{fig:echelle167042} \ref{fig:echelle643} in the appendix). Only the parameters derived directly from \texttt{pySYD} should be considered as the observed global asteroseismic parameters and used in subsequent analyses to calculate stellar mass, radius, and other related properties.

%   \tabletypesize{\scriptsize}
%   \newpage
   \begin{splitdeluxetable*}{lcccccBccccc}
      \tablecaption{Parameters of Host Stars\label{tab:para_all}}
      \tablehead{
         \colhead{Host Name} & \colhead{$T_{eff}$} & \colhead{[Fe/H]} & \colhead{$\nu_{max}$} &\colhead{$\Delta\nu$} & \colhead{$\delta\nu_{02}$} & \colhead{$R_{\star}$} & \colhead{$M_{\star}$} & \colhead{Surface Gravity}  & \colhead{Luminosity} & \colhead{Age} \\
         \colhead{} & \colhead{(K)} & \colhead{(dex)} & \colhead{($\mu Hz$)} & \colhead{($\mu Hz$)} & \colhead{($\mu Hz$)} & \colhead{($R_{\odot}$)} & \colhead{($M_{\odot}$)} & \colhead{($\log_{10}{\mathrm{(cm/s^2)}} $)} & \colhead{($\log_{10}{L_{\odot }} $)} & \colhead{(Gyr)}
      }
      \startdata
      Kepler-278  & 4929$\pm 33$ & 0.290$\pm 0.024$ & $496.841 \pm2.423$  & $30.341\pm0.089$ & $2.965\pm0.016$ & $2.8931\pm0.0269$& $1.2234\pm0.0300$ & $3.6038\pm0.0031$ & $0.6483\pm0.0173$ & $7.010\pm0.581$\\
      HD 167042  & 4926$\pm 79$ & 0.093$\pm 0.061$ & $247.336 \pm1.793$  & $17.172\pm0.049$ & $1.933\pm0.015$ & $4.4947\pm0.0576$ & $1.4696\pm0.0532$   & $3.3008\pm0.0050$  & $1.0301\pm0.0359$ & $7.413\pm0.829$\\
      Kepler-643  & 4812$\pm 41$ & 0.142$\pm 0.040$ & $526.383 \pm3.713$  & $32.990\pm0.308$ & $3.190\pm0.078$ & $2.5615\pm0.0535$ & $1.0039\pm0.0466$ & $3.6237\pm0.0039$ & $0.5008\pm0.0256$ & $10.994\pm1.393$\\
      \enddata
   \end{splitdeluxetable*}

   \section{Calculation of Stellar Parameters} \label{sec:stllr_paras}

   After determining $\nu {max}$ and $\Delta \nu$, along with the $T{eff}$ derived from spectroscopic data, we can estimate the stellar mass, radius, and other physical parameters using asteroseismic scaling relations \citep{1995A&A...293...87K, chaplin2013asteroseismology}. However, for red giants, these scaling relations can be affected by their evolutionary stage and metallicity. \citet{sharma2016stellar} addressed this issue by introducing correction factors into the scaling relations, 

   \begin{equation}\label{equation:1}
      (\frac{R}{R_{\odot } } ) \simeq (\frac{\nu_{max}}{f_{\nu_{max}}\nu_{max,\odot} } )(\frac{ \Delta\nu }{f_{\Delta\nu}\Delta\nu_{\odot} } )^{-2} (\frac{T_{eff}}{T_{eff,\odot} })^{0.5},
   \end{equation}
   \begin{equation}\label{equation:2}
      (\frac{M}{M_{\odot } } ) \simeq (\frac{\nu_{max}}{f_{\nu_{max}}\nu_{max,\odot} } )^{3}(\frac{\Delta\nu}{f_{\Delta\nu}\Delta\nu_{\odot} } )^{-4} (\frac{T_{eff}}{T_{eff,\odot} })^{1.5},
   \end{equation}
   \begin{equation}\label{equation:3}
      \Delta\nu \simeq f_{\Delta\nu}(\frac{\rho}{\rho_{\odot}})^{0.5},
   \end{equation}
   \begin{equation}\label{equation:4}
      \frac{\nu_{max}}{\nu_{max,\odot}} \simeq f_{\nu_{max}}\frac{g}{g_{\odot}}(\frac{T_{eff}}{T_{eff,\odot} })^{-0.5}.
   \end{equation}

   The correction factors, $f_{\nu_{max}}$ and $f_{\Delta\nu}$, are influenced by stellar mass, metallicity, evolutionary stage, and $T_{eff}$ \citep{sharma2016stellar}. The specific correction methods, available through the \texttt{Asfgrid} tool, are detailed on this website\footnote[1]{http://www.physics.usyd.edu.au/k2gap/Asfgrid/} \citep{sharma2016stellar, stello2023extension}. 

   Asteroseismic parameters can also be used to estimate stellar ages. \texttt{PBjam} not only identifies solar-like oscillation modes $\nu_{n,0}$ and $\nu_{n,2}$, but also determines the frequency difference between them, i.e., the small frequency separation $\delta\nu_{02}$. Together with the stellar effective temperature, metallicity, and global asteroseismic parameters derived using \texttt{pySYD}, the stellar age can be calculated based on the seismic scaling relations proposed by \citet{bellinger2019seismic}. The relevant parameters are presented in Table \ref{tab:para_all}.

   \section{Parameter Estimation for Exoplanets Discovered via the Transit Method} \label{sec:transit}

   The two planets in the Kepler-278 system were first discovered by \citet{2014ApJ...784...45R}. Based on their radii, they are classified as Neptune-like planets. However, without radial velocity data, their exact masses cannot be determined. Their host star, Kepler-278, has evolved off the main sequence and is progressing toward the red giant branch.
   Since the radii of the two planets are relatively small compared to the stellar radius, their transit signals are not prominent and require long-cadence data from \textit{Kepler} to obtain high signal-to-noise ratio transit curves. In contrast, the solar-like oscillation signals of the host star, with frequencies in the range of approximately 400 to 600 $\mu Hz$, require short-cadence data for analysis.
   As described in Section \ref{sec:GAP} and \ref{sec:stllr_paras}, we derived parameters such as the mass and radius of Kepler-278 based on its solar-like oscillation data (see Table \ref{tab:para_all}). These parameters will serve as prior distributions for the subsequent calculation of planetary parameters.

   We utilized the Box-Least Squares (BLS, \citet{collier2006fast}) algorithm to search for periodic signals of the two exoplanets. Subsequently, using the Python package \texttt{exoplanet} \citep{exoplanet:joss,exoplanet:zenodo} and its dependencies \citep{exoplanet:arviz,exoplanet:pymc3,exoplanet:theano,exoplanet:kipping13,robitaille2013astropy,exoplanet:astropy18,price2022astropy,exoplanet:luger18,exoplanet:agol20}, we identified the distinct transit curves of the two exoplanets based on their periods and the prior distributions of the host star's mass and radius. The \texttt{exoplanet} Python package is a powerful tool for transit light curve modeling and parameter estimation using Bayesian inference. It integrates Keplerian orbit models and limb darkening effects with PyMC3 \citep{exoplanet:pymc3}, enabling precise Markov Chain Monte Carlo (MCMC) sampling to derive posterior distributions of exoplanetary parameters.

   \begin{deluxetable*}{lcc}[ht!]
      \setlength{\tabcolsep}{8pt}
      \tablecaption{Parameters of exoplanets in Kepler-278 system\label{tab:exo_para1}}
      % \tablewidth{0pt}
      \tablehead{
        \colhead{Parameter} & \colhead{Kepler-278 b} & \colhead{Kepler-278 c}
      }
      \startdata
      $Transit \ Midpoint \ [days]$  & $2454971.520031\pm0.005883$ &  $2454985.353221\pm0.010797$\\
      $P\ [days]$ & $30.1598613\pm0.0001778$ & $51.0773764\pm0.0007119$\\
      $a\ [au]$    & $0.20282\pm0.00166$ & $0.28817\pm0.00236$\\
      $e$   & $0.412\pm0.216$ & $0.287\pm0.181$\\
      $i\ [deg]$   & $86.44\pm1.06$ & $87.94\pm0.46$\\
      $b$   & $0.498\pm0.285$ &  $0.350\pm0.217$\\
      $R_{p} \ [R_{\oplus}]$  & $4.107\pm0.305$  & $3.649\pm0.220$\\
      $R_{p} \ [R_{Jupiter}]$  & $0.3664\pm0.0272$ & $0.3255\pm0.0196$\\
      \enddata
      \tablecomments{In this table, $Transit \ Midpoint$ represents the time of the first transit midpoint in the observational data. $P$ denotes the orbital period of the planet, $a$ is the semi-major axis of the planet's orbit, $e$ represents the orbital eccentricity, $i$ is the orbital inclination, and $b$ is the impact parameter. $R_{p}$ is the planet's radius, expressed in units of Earth radii and Jupiter radii, respectively.}
   \end{deluxetable*}

   \begin{figure*}[ht!]
      \gridline{\fig{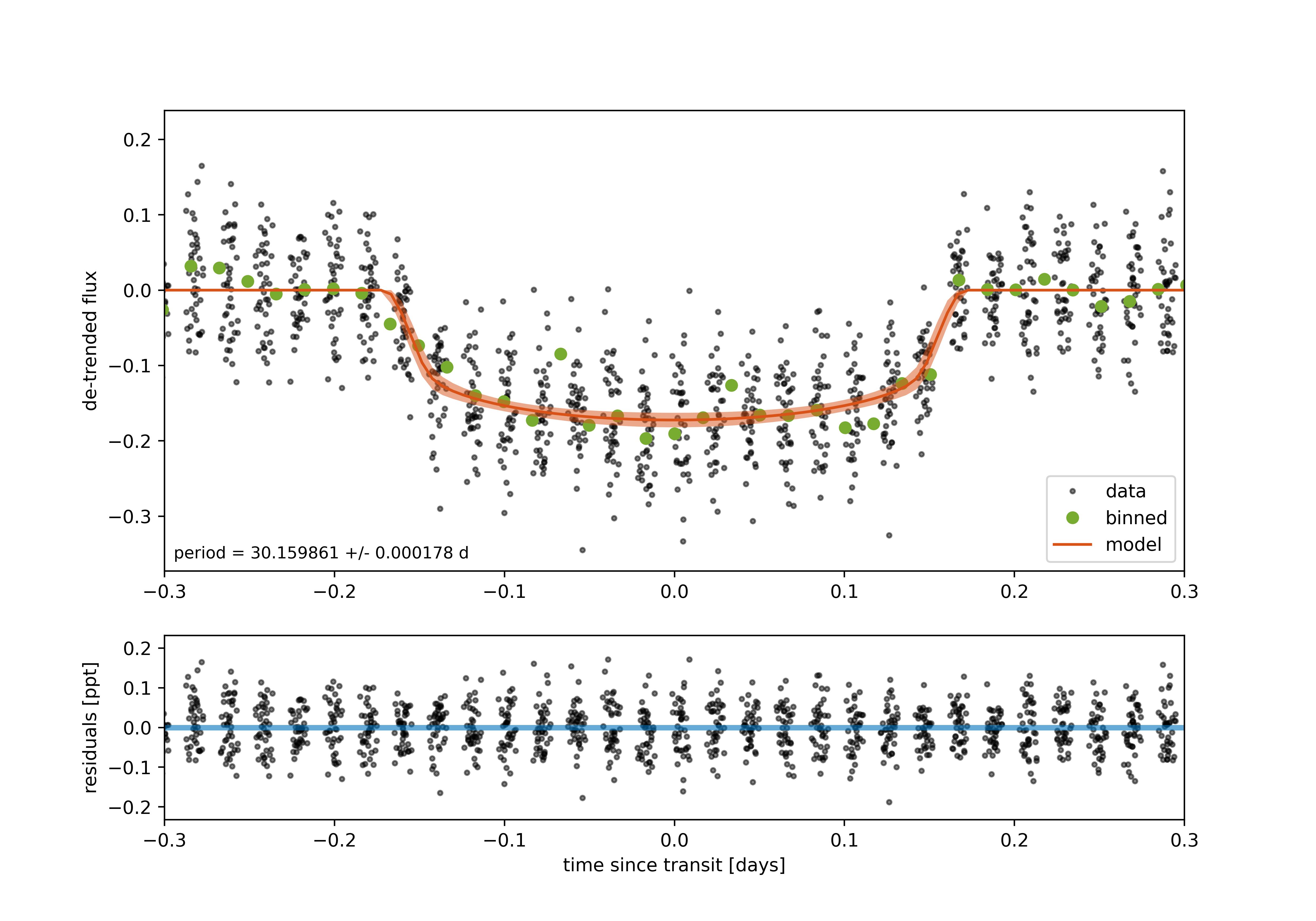}{0.5\textwidth}{Transit Fitting of Kepler-278 b}
      \fig{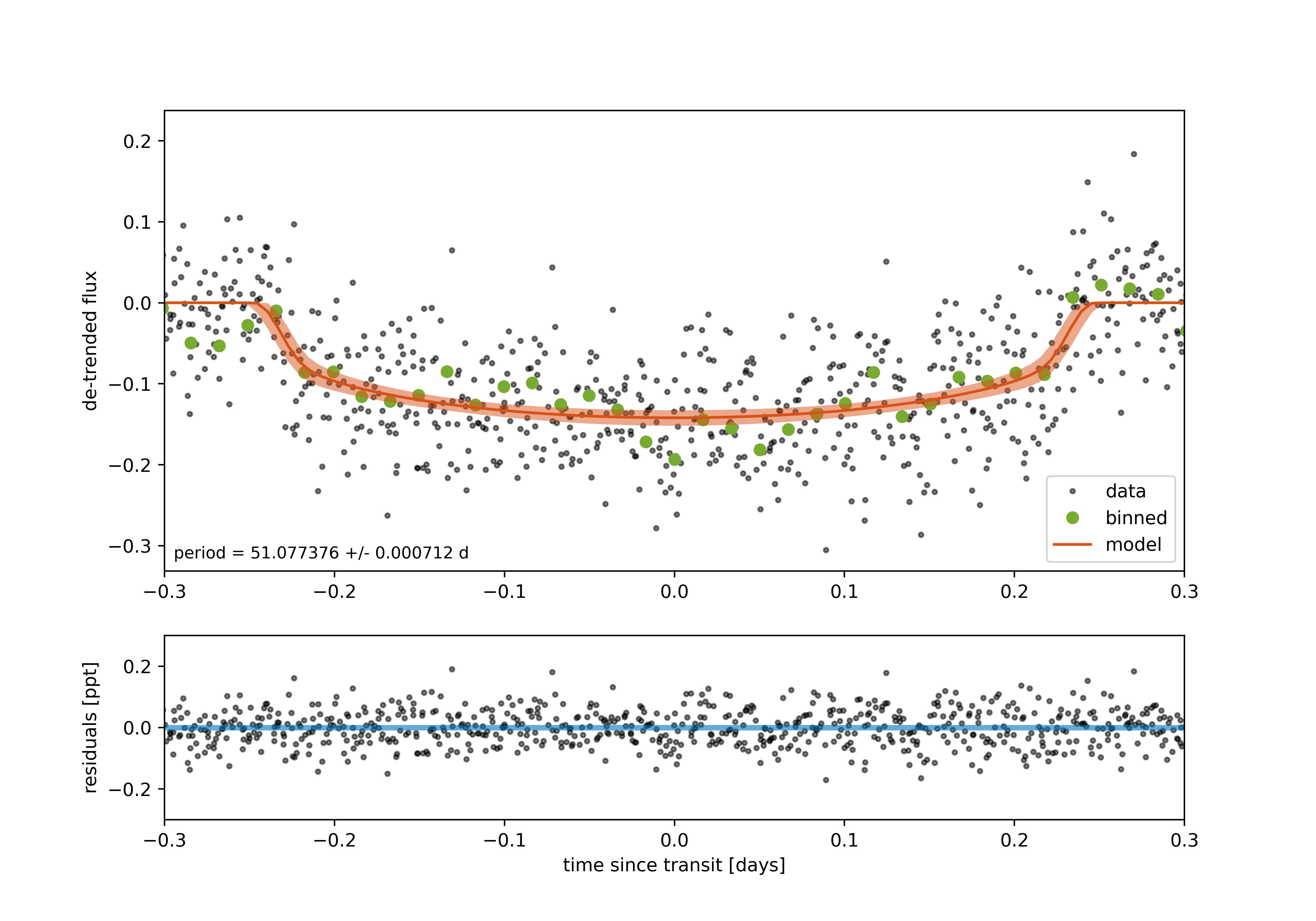}{0.5\textwidth}{Transit Fitting of Kepler-278 c}
      }
      \caption{Transit fitting curves for the two planets in the Kepler-278 system. The black points represent the observed values folded by orbital period, while the red curve indicates the best-fit model. The semi-transparent regions around the red curve correspond to the 1$\sigma$ distribution of the model parameters. It is worth noting that the data points in the left panel appear to be grouped. Upon examining the time axis, it becomes evident that the smallest spacing between these groups matches the observation time resolution, 1800 seconds. This occurs because the orbital period of the planet is approximately an integer multiple of the time resolution.\label{fig:Kepler278transits}}
     \end{figure*}

   The posterior distributions of the parameters for the two planets, calculated using the aforementioned methods, are presented in Table \ref{tab:exo_para1}. The distribution across parameter space is visualized in the corner plots included in the appendix (Section \ref{sec:Appendix}). The results indicate that the planetary parameters derived using stellar parameters obtained from asteroseismology are consistent with those reported in other studies \citep{2014ApJ...784...45R,2016ApJ...822...86M,2018ApJ...866...99B}. This is because the stellar parameters derived from asteroseismology closely align with those obtained through spectroscopic methods, and the parameters of transiting planets are strongly constrained by the mass and radius of their host stars.
   The stellar parameters derived from asteroseismology exhibit noticeably smaller uncertainties compared to those reported in other studies\citep{2014ApJ...784...45R,2016ApJ...822...86M,2018ApJ...866...99B}. However, in this work, the posterior distributions of the planetary parameters show only a modest improvement in uncertainty. We infer that this is due to insufficient photometric precision: the residual plot in Figure \ref{fig:Kepler278transits} reveals significant background noise, which we believe to be the primary source of error in the planetary parameters rather than any issue with the prior distribution of the stellar parameters. Nonetheless, the results demonstrate that using asteroseismology to compute exoplanet parameters is both feasible and reliable.

   \section{Parameter Estimation for Exoplanets Discovered via the Radial Velocity Method} \label{sec:RV}

   HD 167042 b was first discovered by \citet{2008ApJ...675..784J} using the radial velocity method. Since no transit light curve has been detected to date, the orbital inclination of the planet cannot be confirmed. As a result, the radial velocity data can only provide the planet's minimum mass, also known as the mass lower limit ($M_{p}\sin{i}$). This parameter is calculated using the following formula \citep{mayor1995jupiter,cumming2004detectability}:

   \begin{equation}\label{equation:msini}
      M_{p}\sin{i} = \frac{K\sqrt{1-e^{2}} (M_{\star}+M_{p})^{2/3}P^{1/3}}{(2\pi G)^{1/3}} ,
   \end{equation}

   where $K$ is the radial velocity semi-amplitude, $e$ is the orbital eccentricity, $M_{\star}$ is the stellar mass, $M_{p}$ is the planetary mass, $P$ is the orbital period, and $G$ is the gravitational constant. In the formula, calculating the planet's minimum mass requires the sum of the stellar and planetary masses. In practice, the stellar mass is significantly greater than the planetary mass, so this term is approximated using only the stellar mass. Other parameters, such as $K$ (the radial velocity semi-amplitude), $P$ (the orbital period), and $e$ (the orbital eccentricity), are obtained from fitting the radial velocity curve.
   Since no additional radial velocity data were observed in this study, we adopted the results from \citet{2023PASJ...75.1030T}, as shown in Table \ref{tab:exo_para2}. The stellar mass, determined through asteroseismic analysis in this study, directly constrains the planet's minimum mass. Additionally, the semi-major axis of the planet's orbit is derived from the stellar mass and the parameters fitted from the radial velocity data, using the following formula:
   
   \begin{equation}\label{equation:msini}
      a = (\frac{GM_{\star}P^{2}}{4\pi ^{2}})^{1/3} ,
   \end{equation}
   
   where $a$ is the semi-major axis, $M_{\star}$ is the stellar mass, $P$ is the orbital period, and $G$ is the gravitational constant. It can be seen that for exoplanets discovered via the radial velocity method, their two key parameters—the minimum mass and the semi-major axis—are directly determined by the stellar mass and are independent of the stellar radius. This characteristic distinguishes them from exoplanets discovered via the transit method (see Section \ref{sec:transit}).

   \begin{deluxetable*}{lc}[ht!]
      \setlength{\tabcolsep}{8pt}
      \tablecaption{Parameters of the exoplanet in HD 167042 system\label{tab:exo_para2}}
      % \tablewidth{0pt}
      \tablehead{
        \colhead{Parameter} & \colhead{HD 167042 b} 
      }
      \startdata
      $P^{\dagger} \ [days]$ & $417.67^{+0.43}_{-0.41}$ \\
      $K^{\dagger}$ [$m*s^{-1}$] & $30.52^{+0.70}_{-0.95}$  \\
      $e^{\dagger}$   & $0.010^{+0.033}_{-0.004}$ \\
      $a \ [au]$    & $1.2432\pm0.0151$ \\
      $M_{p}\sin{i} \ [M_{\oplus}]$   & $460.98\pm15.35$ \\
      $M_{p}\sin{i} \ [M_{Jupiter}]$   & $1.4504\pm0.0483$ \\
      \enddata
      \tablecomments{In this table, $P$ represents the orbital period of the planet, $K$ denotes the amplitude of the radial velocity variation of the host star, $e$ is the orbital eccentricity of the planet, $a$ is the semi-major axis of the orbit, and $M_{p}\sin{i}$ indicates the minimum mass of the planet, expressed in both Earth masses and Jupiter masses. Since the orbital inclination of the planet cannot currently be measured, the exact mass value cannot be determined.
      The three values marked with $\dagger$ are referenced from the work of \citet{2023PASJ...75.1030T}, while the subsequent parameter values are derived from the calculations in this study.}
   \end{deluxetable*}

   Compared with the work of \citet{2023PASJ...75.1030T}, our stellar mass derived from asteroseismology is slightly larger (but still within 1 $\sigma$), which consequently increases both the minimum planet mass and the orbital semi-major axis. However, there are some interesting findings regarding parameter precision. The uncertainty in the planet's minimum mass in our study is close to, though slightly larger than, that reported by \citet{2023PASJ...75.1030T}, which is not unexpected. For the orbital semi-major axis, our uncertainty is 0.0151 $au$, whereas \citet{rosenthal2021california} report 0.042 $au$ and \citet{bowler2009retired} report 0.04 $au$. In contrast, the uncertainty from \citet{2023PASJ...75.1030T}. is notably smaller, at only 0.001 $au$.

   \section{Parameter Estimation for Exoplanets with Both Radial Velocity and Transit Data} \label{sec:trans_and_RV}

   Exoplanets with only transit light curves cannot determine the planet's mass, while those with only radial velocity data cannot determine the planet's radius. Only by combining transit light curves and radial velocity data can we obtain comprehensive and precise planetary parameters, with Kepler-643 b serving as a prime example.
   Kepler-643 b was first discovered by \citet{2016ApJS..225....9H} using transit data from \textit{Kepler}. Later, in 2018, \citet{2018ApJ...861L...5G} provided radial velocity measurements, enabling the determination of the planet's exact mass. This planet is a gas giant with both its mass and radius closely resembling those of Jupiter.

   The process of analyzing the light curve of this planet is highly similar to that of planets in Kepler-278 system. The long-cadence (30-minute) data from \textit{Kepler} provides higher signal-to-noise ratio transit curves, while the short-cadence (1-minute) data enables the detection of higher-frequency solar-like oscillation signals from the host star. The distinct frequency domains of the transit signals and asteroseismic signals in the power spectrum ensure that the two types of signals do not interfere with each other. We used the BLS algorithm to identify the orbital period of the planet. Stellar parameters obtained through asteroseismology were applied as prior distributions. Using the exoplanet package, we constructed a transit model for the exoplanet. Unlike the analysis in Section \ref{sec:transit}, we additionally included a radial velocity model using the exoplanet package. The radial velocity data used for the fitting were sourced from \citet{2018ApJ...861L...5G}. Finally, MCMC sampling was performed to obtain the posterior distributions of the planetary parameters.

   \begin{figure*}[ht!]
      \gridline{\fig{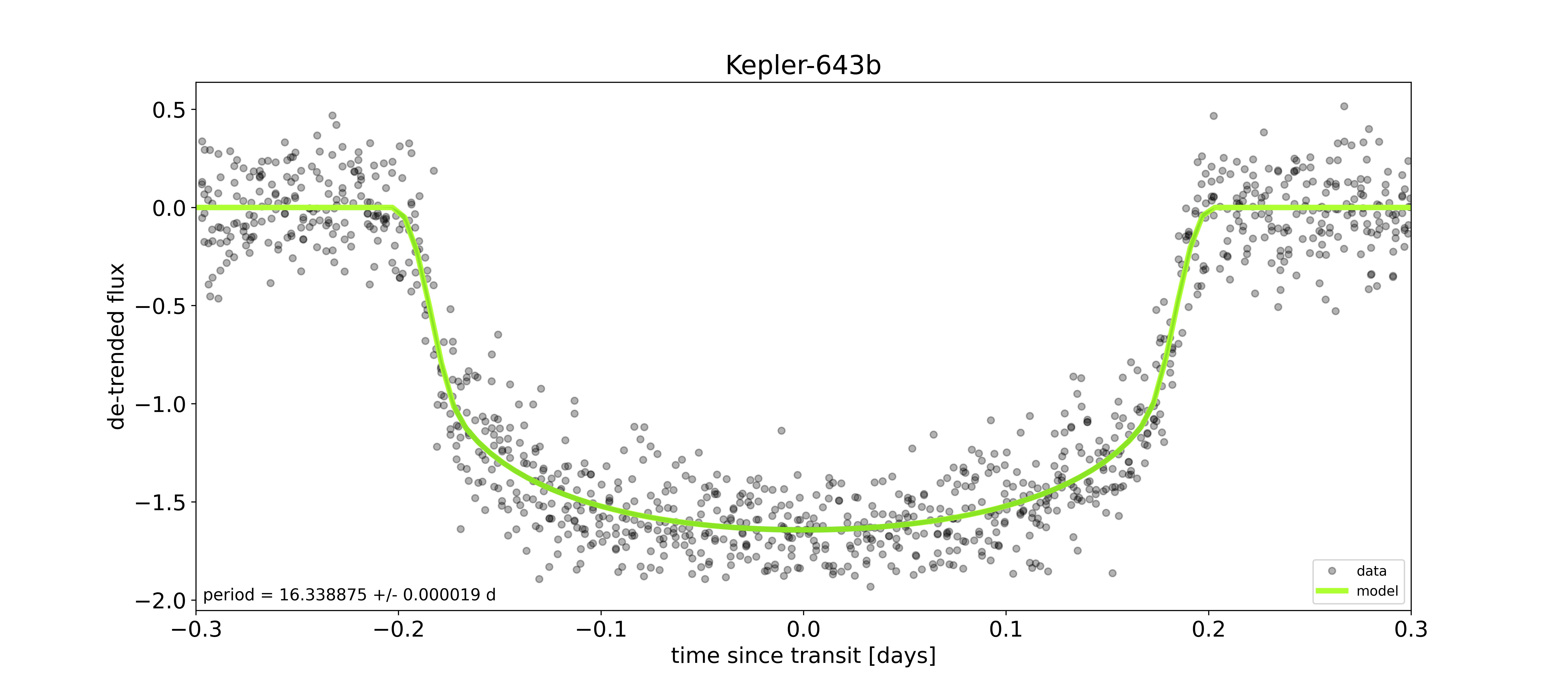}{0.5\textwidth}{Transit Fitting of Kepler-643 b}
      \fig{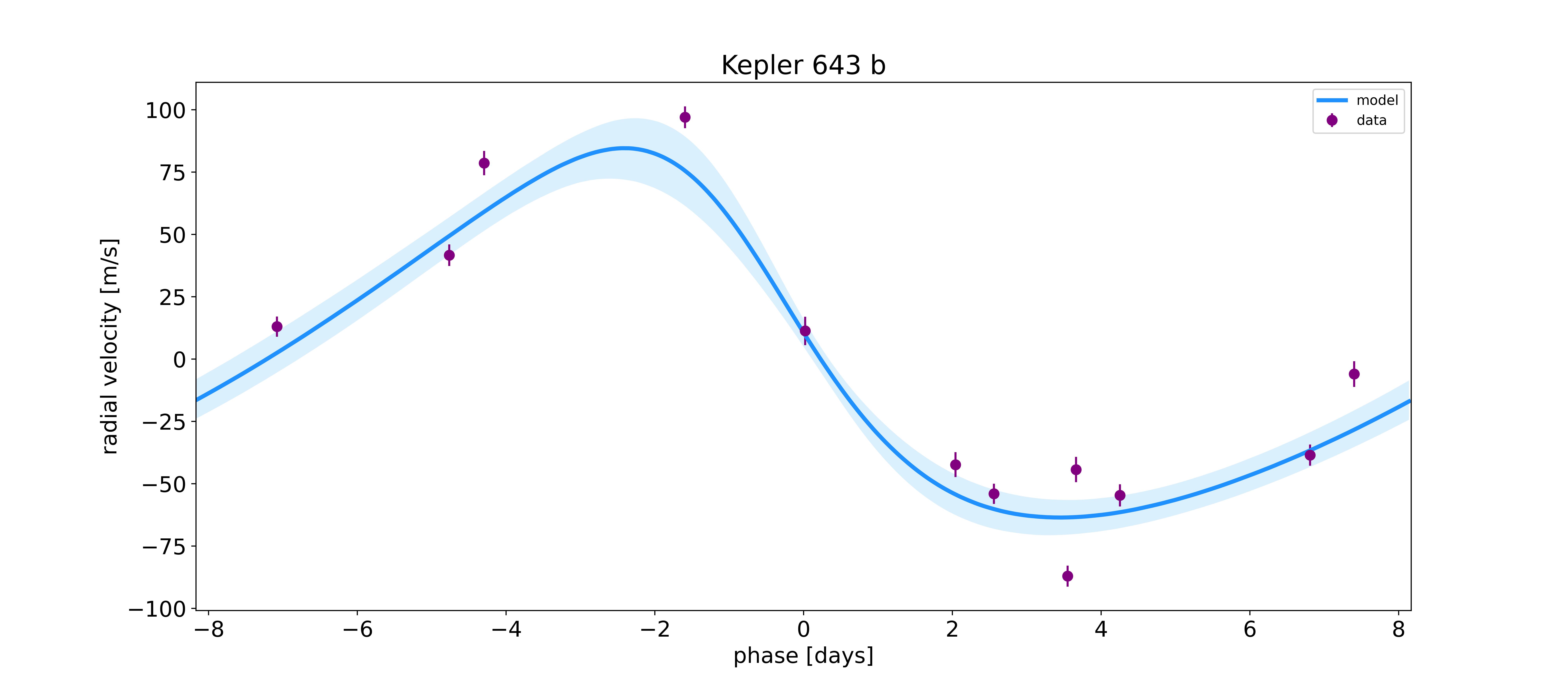}{0.5\textwidth}{RV Fitting of Kepler-643 b}
      }
      \caption{Transit and radial velocity fitting plots for Kepler-643 b. In the left panel, the black points represent the observed data folded by the orbital period of the planet, while the green curve shows the best-fit model. Although the 1$\sigma$ distribution of the model parameters is also present in the figure, it is entirely covered by the green curve and therefore not visible. In the right panel, the purple points with error bars represent the radial velocity observations, as reported in the work of \citet{2018ApJ...861L...5G}. The blue curve indicates the best-fit model, while the semi-transparent blue region corresponds to the 1$\sigma$ distribution of the model parameters.}
     \end{figure*}

     \begin{deluxetable*}{lc}[ht!]
      \setlength{\tabcolsep}{8pt}
      \tablecaption{Parameters of the exoplanet in Kepler-643 system\label{tab:exo_para3}}
      % \tablewidth{0pt}
      \tablehead{
        \colhead{Parameter} & \colhead{Kepler-643 b} 
      }
      \startdata
      $Transit \ Midpoint \ [days]$  & $2455010.918065\pm0.000938$ \\
      $P \ [days]$ & $16.33887536\pm0.00001866$ \\
      $e$   &  $0.263\pm0.053$\\
      $a \ [au]$ & $0.1262\pm0.0019$   \\
      $i\ [deg]$ & $88.15\pm0.99$ \\
      $b$   & $0.502\pm0.288$ \\
      $\omega \ [deg]$ &  $55.71\pm16.59$ \\
      $R_{p} \ [R_{\oplus}]$   &  $10.382\pm0.268$ \\
      $R_{p} \ [R_{Jupiter}]$   &  $0.9263\pm0.0239$\\
      $M_{p} \ [M_{\oplus}]$   &  $286.08\pm34.56$\\
      $M_{p} \ [M_{Jupiter}]$   & $0.900\pm0.108$\\
      \enddata
      \tablecomments{The first six parameters in this table have the same meanings as those in Table \ref{tab:exo_para1} and will not be reiterated here. Among the subsequent parameters, $\omega$ represents the argument of periastron, and $M_{p}$ denotes the mass of the planet, expressed in both Earth masses and Jupiter masses.}
   \end{deluxetable*}

   Compared to the analysis of planets with only transit data, incorporating radial velocity (RV) data allows for the determination of additional and more precise parameters, most notably the planet's mass and orbital eccentricity. Radial velocity measurements can confirm the planet's minimum mass ($M_{p}\sin{i}$), while transit data provide the orbital inclination. Combining these results enables the determination of the planet's true mass. Measuring the orbital eccentricity directly from the transit curve is challenging unless the transit light curve has an extremely high signal-to-noise ratio. Otherwise, the posterior distribution from MCMC sampling may deviate significantly from a Gaussian distribution. In contrast, eccentricity is directly reflected in the shape of the RV curve. Therefore, including RV data in the analysis yields a much more accurate estimate of the orbital eccentricity. 

   The key parameters of Kepler-643 b are listed in Table \ref{tab:exo_para3}. Compared to \citet{2018ApJ...861L...5G}, the stellar parameters we derived from asteroseismology are slightly smaller (within 2 $\sigma$), resulting in correspondingly smaller planetary radius and mass. However, because the uncertainties in our stellar parameters are smaller, the uncertainties in the posterior distributions of the planetary parameters are also reduced. On the other hand, our planetary parameter results are nearly identical to those of \citet{2016ApJ...822...86M}, since their stellar parameters and uncertainties closely match ours. Overall, our work provides more comprehensive parameters with higher precision than previous studies. In summary, these findings highlight that the determination of planetary parameters critically depends on the properties of the host star. Asteroseismology, as a discipline dedicated to unveiling stellar characteristics, plays a vital role in indirectly measuring planetary parameters, underscoring its significant reference value.

   \section{Conclusion} \label{sec:conclusion}

   This study presents a comprehensive framework for determining the physical parameters of host stars and their orbiting exoplanets through a combination of asteroseismic analysis and spectroscopic observations. By analyzing light curves from \textit{Kepler} and \textit{TESS}, we derived the global asteroseismic parameters $\nu_{\text{max}}$ and $\Delta \nu$ for three host stars: Kepler-278, HD 167042, and Kepler-643. The power spectra were carefully analyzed to identify solar-like oscillation modes. Using the results from the \texttt{pySYD} and \texttt{PBjam} tools, combined with spectroscopic effective temperatures, we successfully determined precise stellar masses, radii, and other fundamental parameters. For example, Kepler-278 was found to have a radius of $2.8931 \pm 0.0269 , R_{\odot}$ and a mass of $1.2234 \pm 0.0300 , M_{\odot}$, while HD 167042 exhibited a larger radius of $4.4947 \pm 0.0576 , R_{\odot}$ and a mass of $1.4696 \pm 0.0532 , M_{\odot}$.

   The derived stellar parameters played a pivotal role in estimating the physical properties of their orbiting exoplanets. Although the planetary parameters were not directly observed, the indirect constraints based on host star properties proved to be reliable and efficient. For instance, the planetary radii of Kepler-643 b were estimated with an uncertainty of less than 3\%, demonstrating the robustness of the combined asteroseismic and spectroscopic approach. The ability to infer planetary properties indirectly through host star characterization provides a practical solution for cases where direct measurements are not feasible, especially for systems observed by missions such as \textit{Kepler} and \textit{TESS}.

   In addition to showcasing the precision of asteroseismic methods, this study emphasizes their versatility in studying a wide range of stellar and planetary systems. The detailed characterization of subgiant and red giant host stars offers valuable insights into the physical conditions under which planets orbit stars in advanced evolutionary stages. This work not only underscores the transformative potential of asteroseismology in exoplanet research but also demonstrates its applicability across diverse stellar populations.

   In summary, the results presented here highlight the critical role of asteroseismic analysis in improving the accuracy of both stellar and planetary parameter estimation. By combining observational data with theoretical models, this work provides a solid foundation for future studies aiming to unravel the complex dynamics of planetary systems. The methodologies and findings detailed in this paper will undoubtedly serve as a cornerstone for advancing exoplanetary science in the era of next-generation space missions like PLATO and JWST.

\section*{Acknowledgments}

	This work is supported by National Key R\&D Progrom of China (grant No:2022YFE0116800),the National Natural Science Foundation of China (No. 11933008 and No. 12103084), the basic research project of Yunnan Province (Grant No. 202301AT070352). This paper includes data collected by the \textit{Kepler} and \textit{TESS} mission, which are publicly available from the Mikulski Archive for Space Telescopes (MAST)\footnote{https://mast.stsci.edu/}. The data used in this work can be found at this \dataset[link]{http://dx.doi.org/10.17909/jhv7-w834}.

\bibliography{sample631}{}
\bibliographystyle{aasjournal}

\newpage
\section{Appendix}\label{sec:Appendix}
	This is a supplementary materials to the paper, and it contains 1 tables (Table \ref{tab:nu0,nu2}) and 12 figures (Figure \ref{fig:corner278b}, \ref{fig:echelle278}, \ref{fig:pysyd_corner_278}, \ref{fig:corner278c}, \ref{fig:echelle167042}, \ref{fig:pysyd_corner_167042}, \ref{fig:echelle643}, \ref{fig:pysyd_corner_643}, \ref{fig:corner643b}).
   In the Appendix, we present the power spectrum analysis of the host stars performed using \texttt{PBjam}, along with three echelle diagrams that illustrate their solar-like oscillation modes in a more intuitive manner. As shown in these diagrams, the width, known as $\Delta\nu$, aligns with the value derived from \texttt{pySYD}, confirming the reliability of the \texttt{pySYD} results. However, there are minor differences in the numerical values because \texttt{pySYD} provides the observed  $\Delta\nu$, which is the value ultimately used in calculating the stellar parameters.

   \renewcommand\thefigure{A}
   \begin{figure*}[ht!]
      \includegraphics[width=0.95\linewidth]{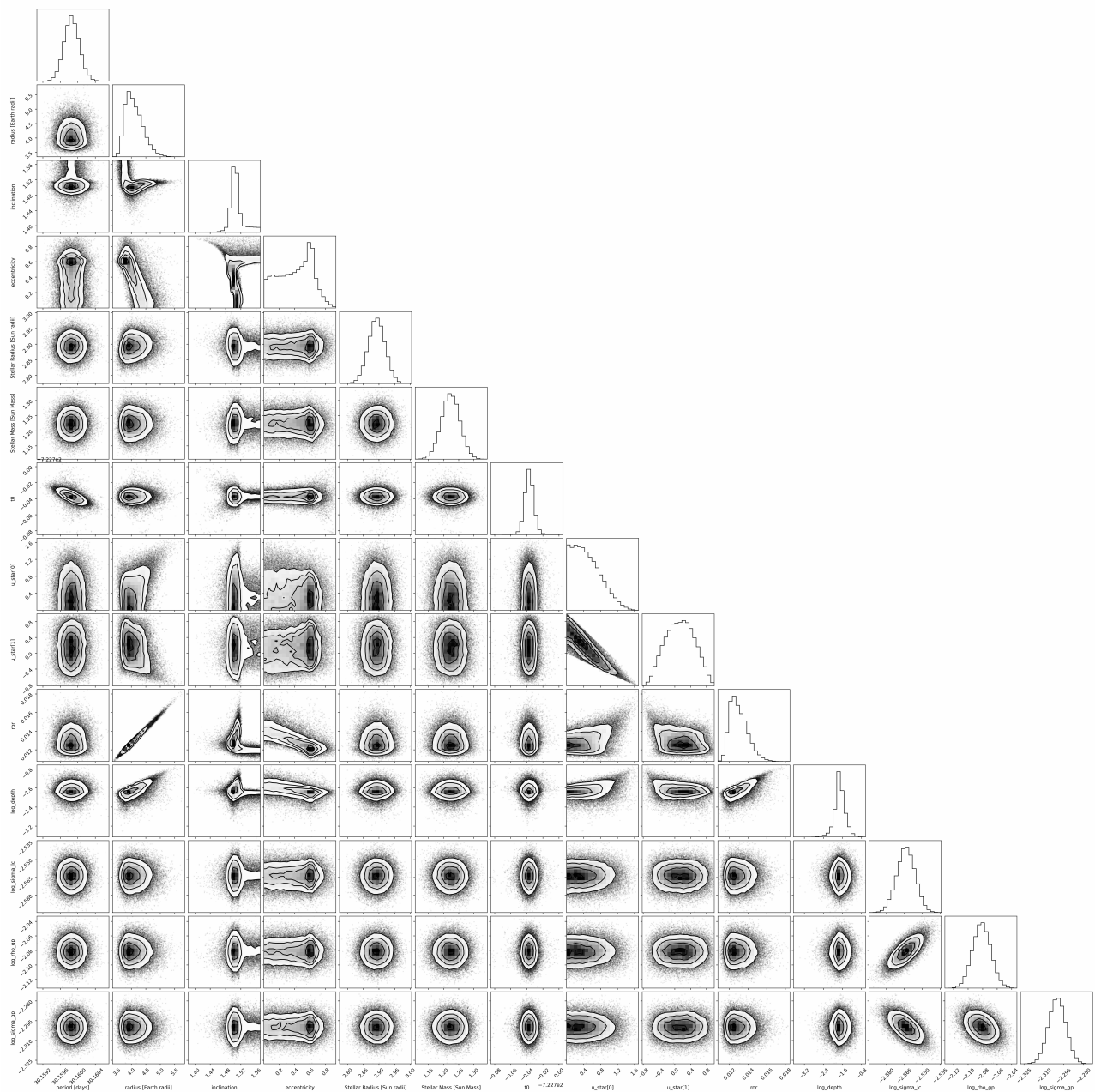}
      \centering
      \caption{Corner diagram of the parameter fitting for Kepler-278 b \label{fig:corner278b}}
   \end{figure*}

   \renewcommand\thefigure{B}
   \begin{figure*}[ht!]
      \includegraphics[width=0.95\linewidth]{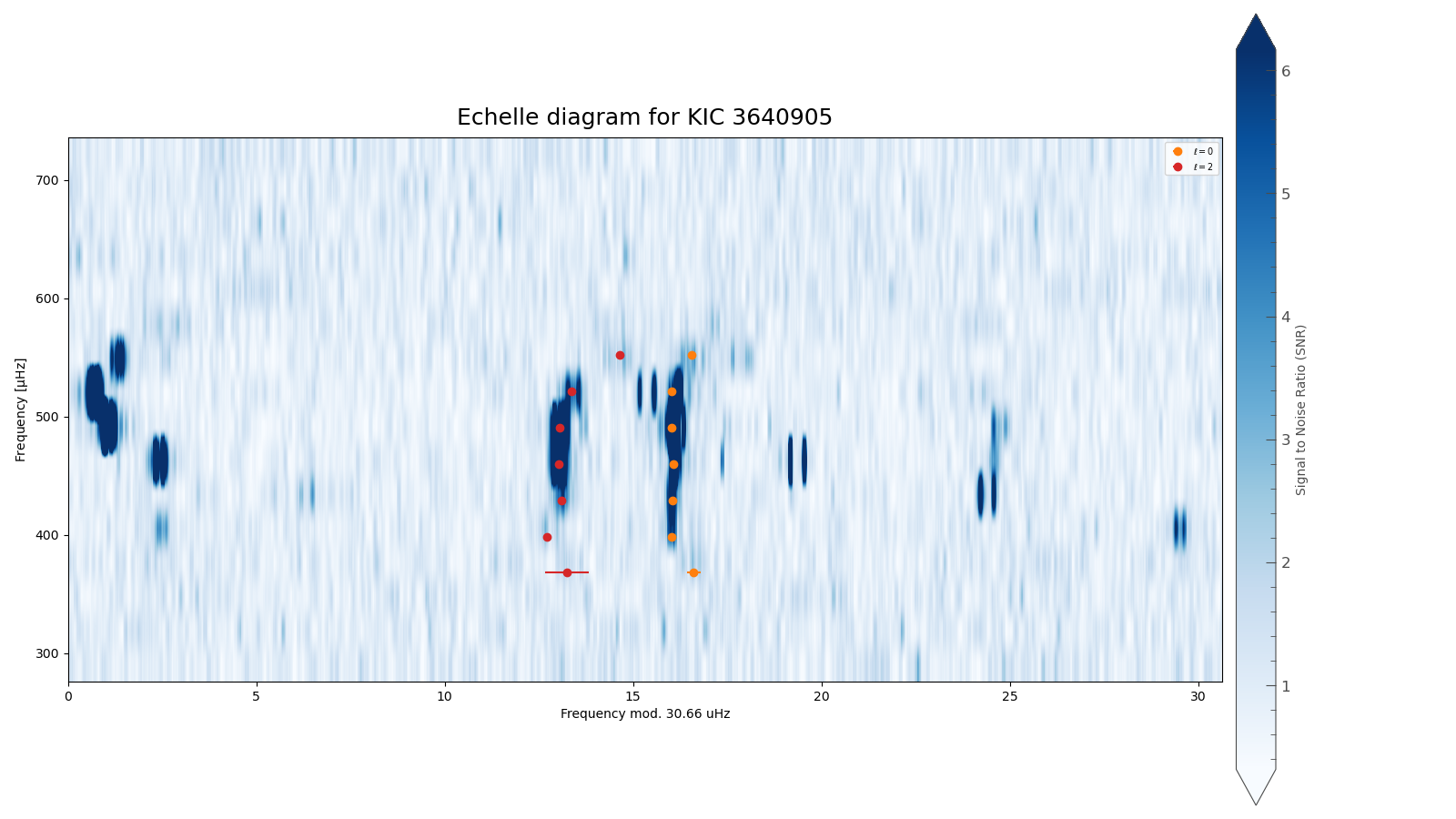}
      \centering
      \caption{The echelle diagrams of host stars Kepler-278, the orange and red dots represent $\nu_{n,0}$ and $\nu_{n,2}$ identified by \texttt{PBjam}.\label{fig:echelle278}}
   \end{figure*}

   \renewcommand\thetable{A}
   \begin{deluxetable*}{ccccccc}
   \tablecaption{$\nu_{n,0}$ and $\nu_{n,2}$ of Host Stars identified by \texttt{PBjam}
   \label{tab:nu0,nu2}}
   \tablehead{
      \colhead{Host Name} & 
      \multicolumn{2}{c}{Kepler-278} & 
      \multicolumn{2}{c}{HD 167042} & 
      \multicolumn{2}{c}{Kepler-643} \\[2pt]
      \colhead{Row} &
      \colhead{$l=0$} &
      \colhead{$l=2$} &
      \colhead{$l=0$} &
      \colhead{$l=2$} &
      \colhead{$l=0$} &
      \colhead{$l=2$}
   }
   \startdata
   1 & $384.560 \pm 0.188$ & $381.194 \pm 0.569$ & $179.257 \pm 0.062$ & $176.454 \pm 0.188$ & $418.029 \pm 0.883$ & $414.449 \pm 0.998$ \\
   2 & $414.634 \pm 0.032$ & $411.333 \pm 0.090$ & $195.860 \pm 0.022$ & $193.716 \pm 0.111$ & $450.163 \pm 0.264$ & $446.656 \pm 0.490$ \\
   3 & $445.312 \pm 0.017$ & $442.378 \pm 0.036$ & $212.818 \pm 0.015$ & $210.744 \pm 0.042$ & $483.927 \pm 0.065$ & $481.284 \pm 0.286$ \\
   4 & $476.015 \pm 0.012$ & $472.966 \pm 0.020$ & $230.008 \pm 0.014$ & $228.068 \pm 0.042$ & $517.184 \pm 0.021$ & $514.493 \pm 0.052$ \\
   5 & $506.633 \pm 0.014$ & $503.660 \pm 0.020$ & $247.070 \pm 0.010$ & $245.120 \pm 0.026$ & $550.524 \pm 0.037$ & $546.545 \pm 0.047$ \\
   6 & $537.282 \pm 0.046$ & $534.634 \pm 0.032$ & $264.308 \pm 0.015$ & $262.483 \pm 0.036$ & $584.209 \pm 0.041$ & $580.872 \pm 0.068$ \\
   7 & $568.491 \pm 0.098$ & $566.579 \pm 0.099$ & $281.756 \pm 0.025$ & $279.795 \pm 0.031$ & \nodata           & \nodata           \\
   8 & \nodata           & \nodata           & $299.070 \pm 0.048$ & $297.105 \pm 0.051$ & \nodata           & \nodata           \\
   \enddata
   \tablecomments{
      Row numbers (1, 2, 3, …) are for reference only and do not correspond to the radial order $n$. 
      Missing entries (\nodata) indicate frequencies not identified or not applicable in the dataset.
   }
   \end{deluxetable*}

   \renewcommand\thefigure{C}
   \begin{figure*}[ht!]
    \gridline{\fig{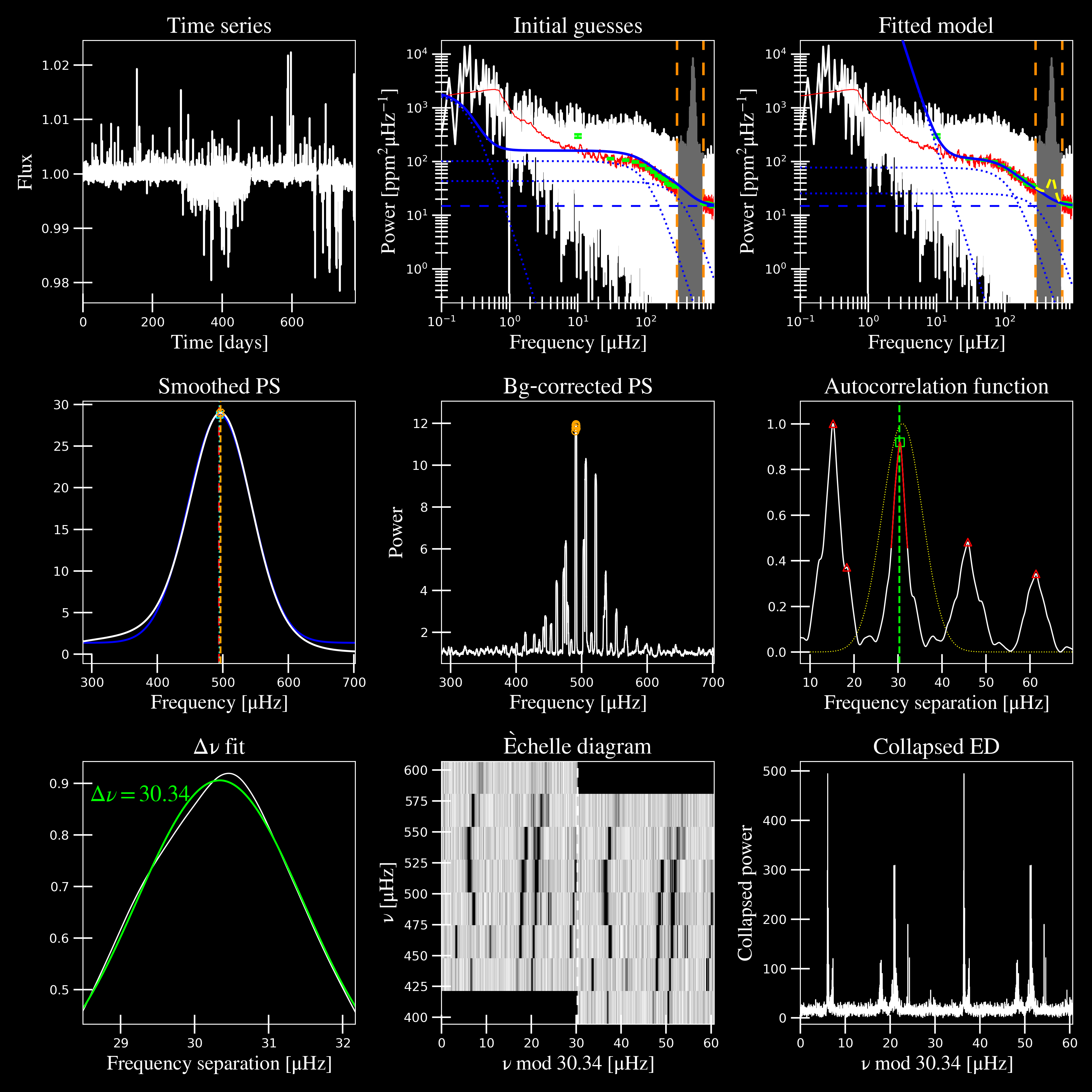}{0.5\textwidth}{Asteroseismic Parameter Outputs from \texttt{pySYD}}
    }
    \gridline{\fig{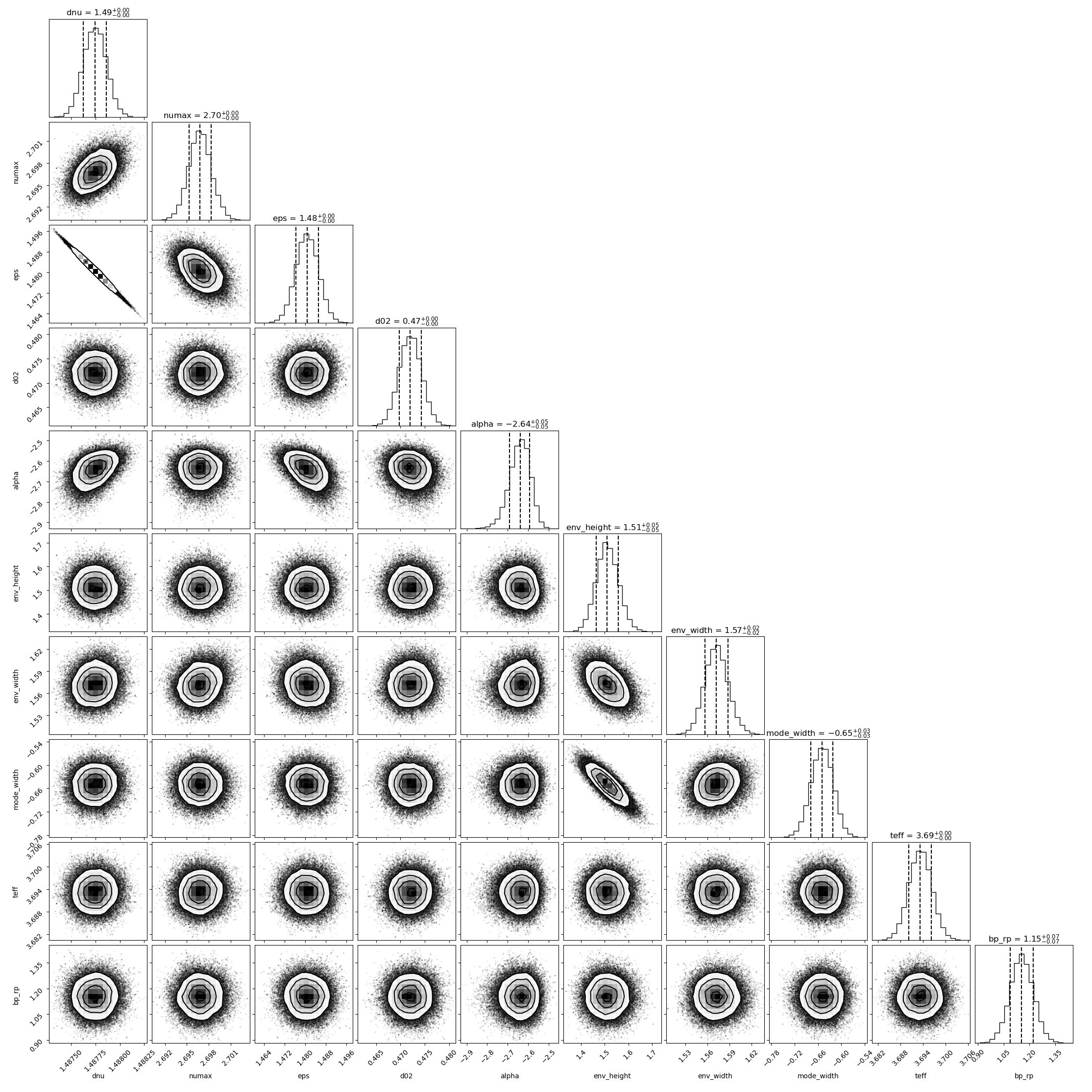}{0.7\textwidth}{Corner diagram of asymptotic fit parameters}
    }
    \caption{Kepler-278 \label{fig:pysyd_corner_278}}
   \end{figure*}

   \renewcommand\thefigure{D}
   \begin{figure*}[ht!]
      \includegraphics[width=0.95\linewidth]{corner-Kepler278c.jpg}
      \centering
      \caption{Corner diagram of the parameter fitting for Kepler-278 c \label{fig:corner278c}}
   \end{figure*}

   \renewcommand\thefigure{E}
   \begin{figure*}[ht!]
      \includegraphics[width=0.95\linewidth]{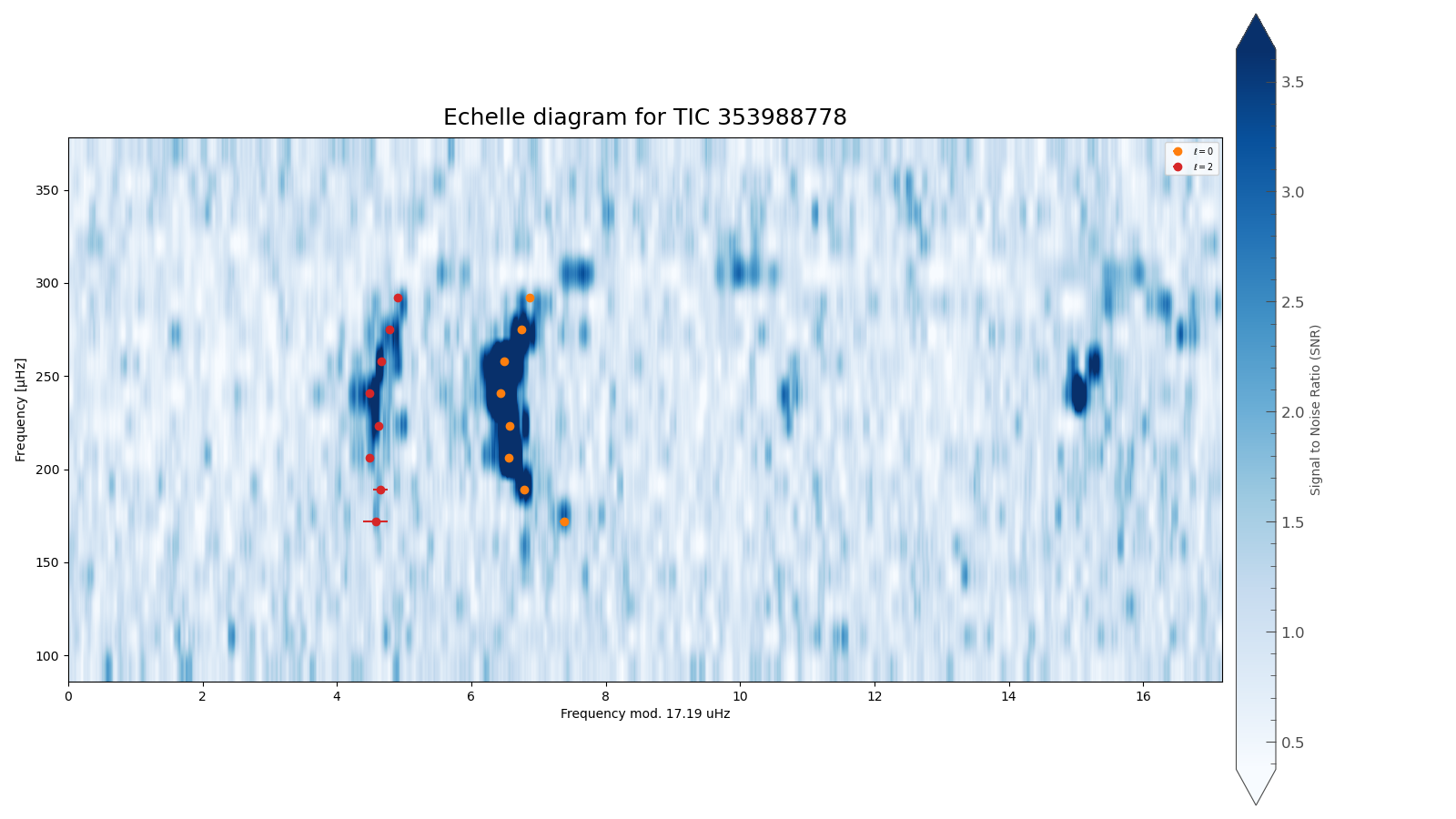}
      \centering
      \caption{The echelle diagrams of host stars HD 167042, the orange and red dots represent $\nu_{n,0}$ and $\nu_{n,2}$ identified by \texttt{PBjam}.\label{fig:echelle167042}}
   \end{figure*}

   \renewcommand\thefigure{F}
   \begin{figure*}[ht!]
    \gridline{\fig{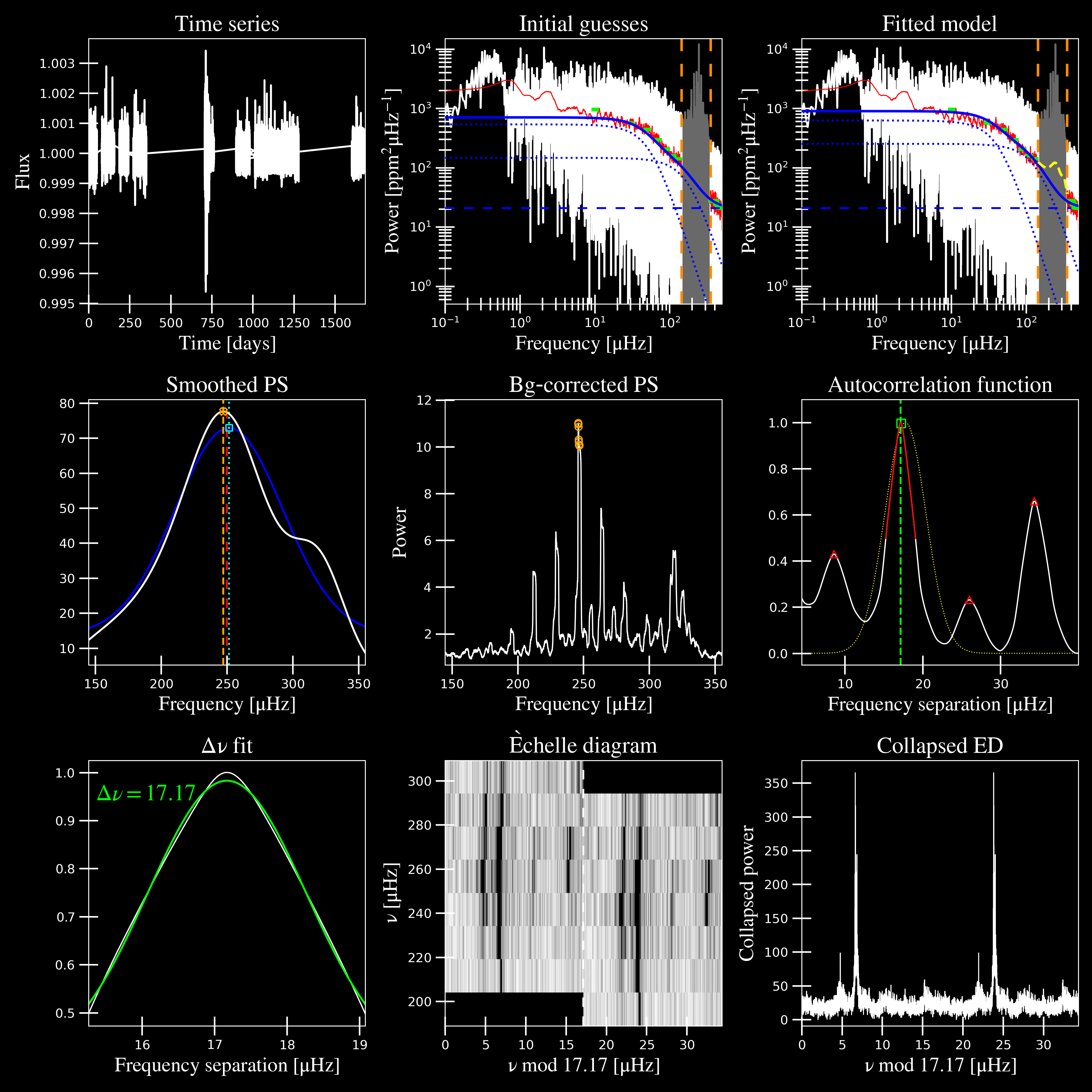}{0.5\textwidth}{Asteroseismic Parameter Outputs from \texttt{pySYD}}
    }
    \gridline{\fig{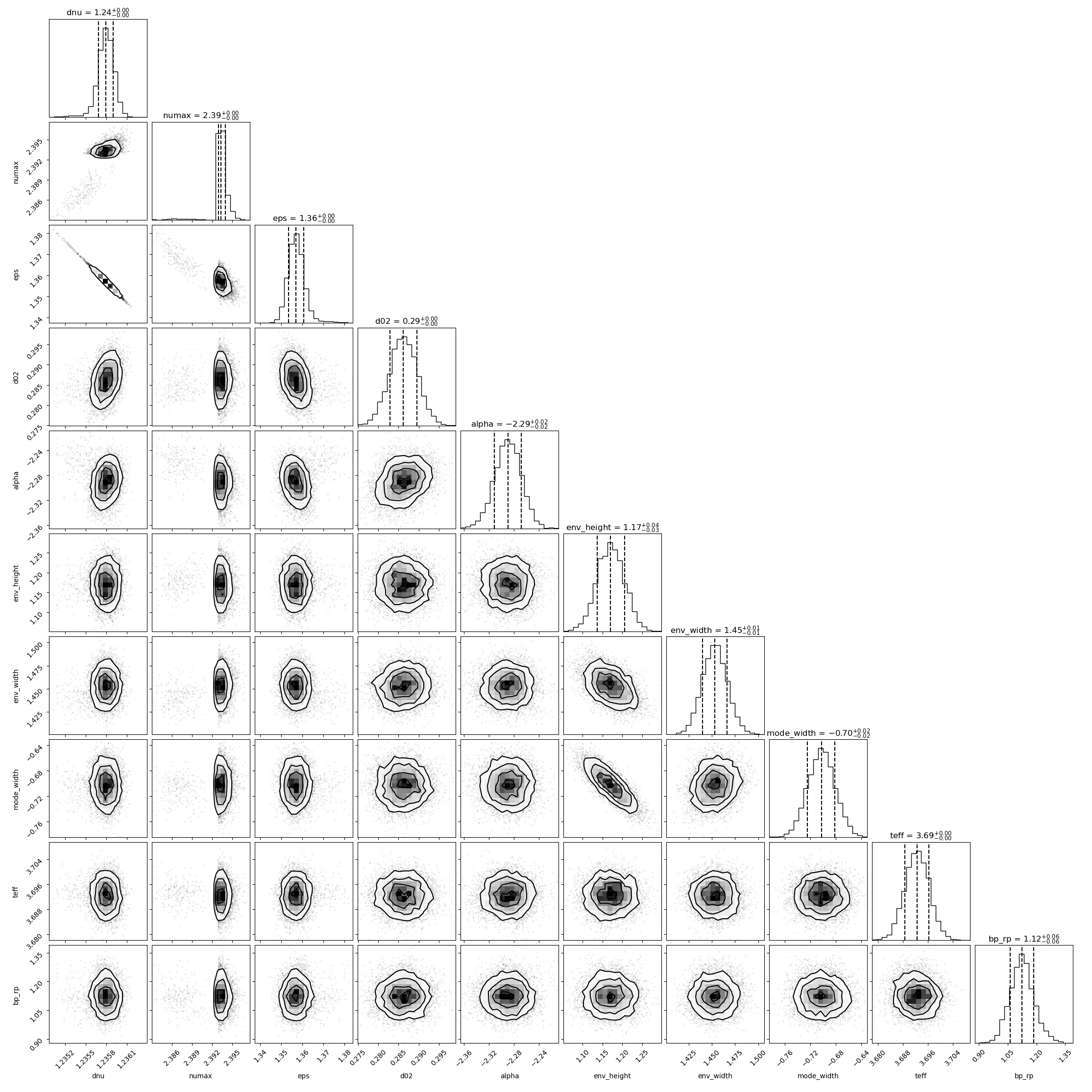}{0.7\textwidth}{Corner diagram of asymptotic fit parameters}
    }
    \caption{HD 167042 \label{fig:pysyd_corner_167042}}
   \end{figure*}

   \renewcommand\thefigure{G}
   \begin{figure*}[ht!]
      \includegraphics[width=0.95\linewidth]{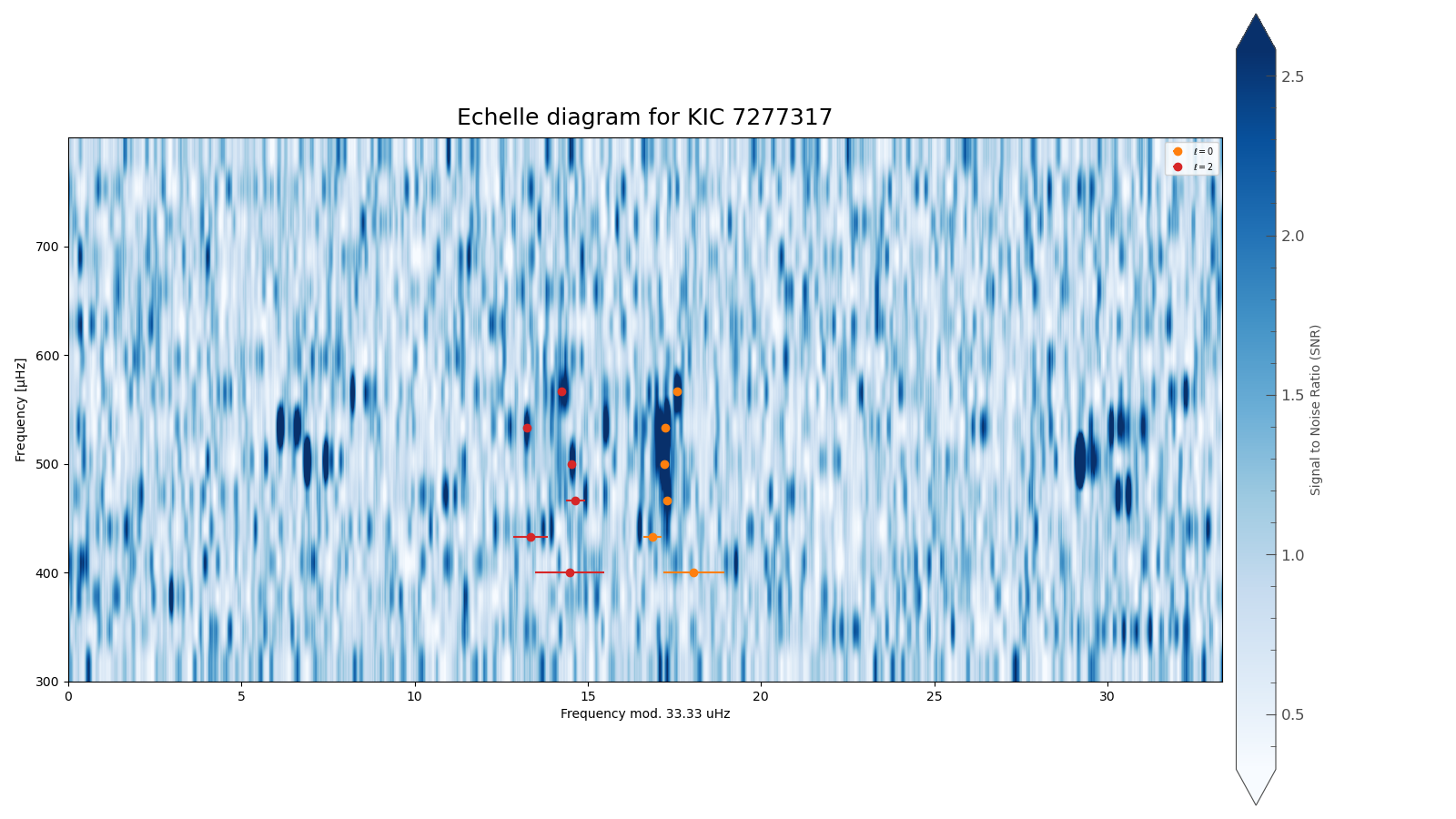}
      \centering
      \caption{The echelle diagrams of host stars Kpeler-643, the orange and red dots represent $\nu_{n,0}$ and $\nu_{n,2}$ identified by \texttt{PBjam}.\label{fig:echelle643}}
   \end{figure*}

   \renewcommand\thefigure{H}
   \begin{figure*}[ht!]
    \gridline{\fig{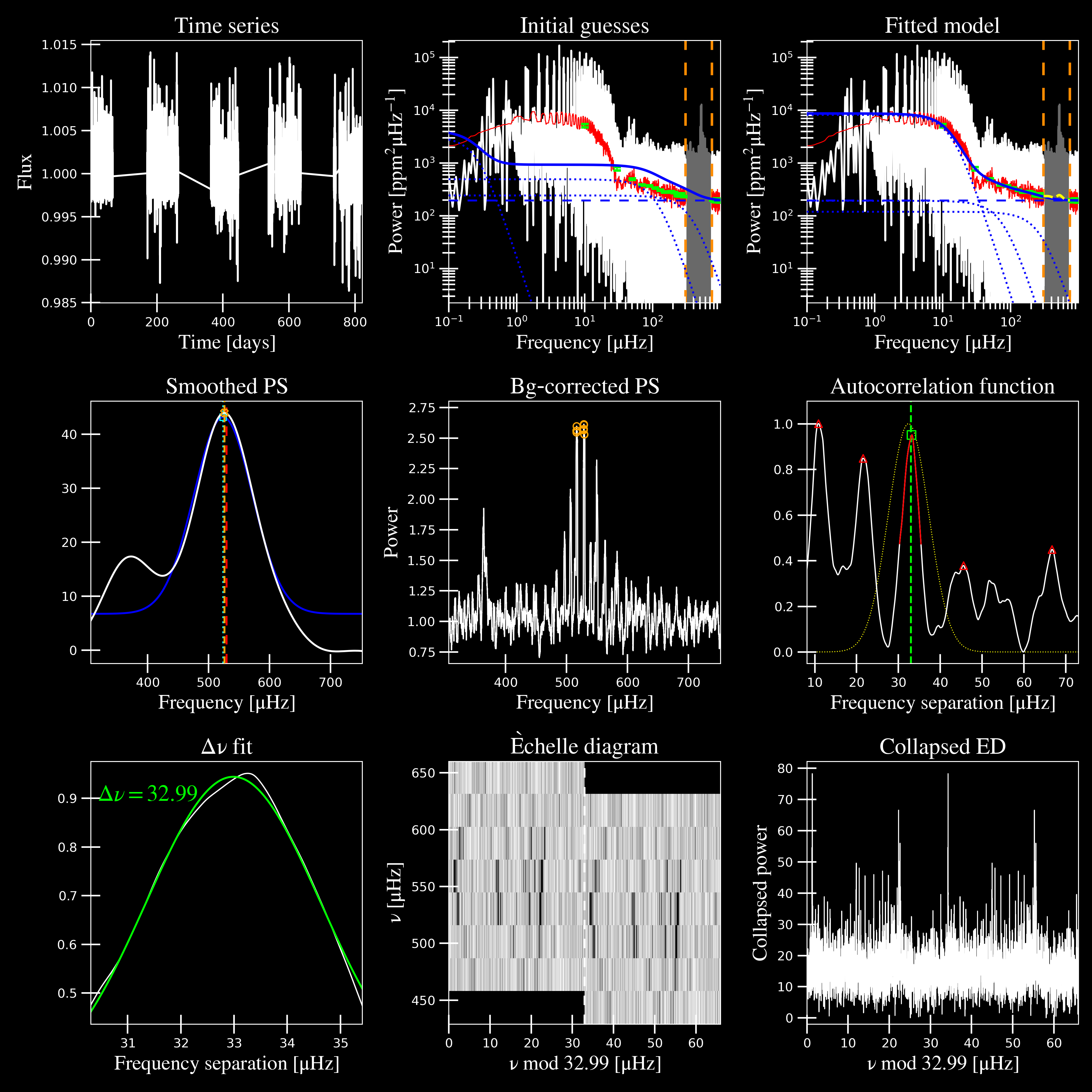}{0.5\textwidth}{Asteroseismic Parameter Outputs from \texttt{pySYD}}
    }
    \gridline{\fig{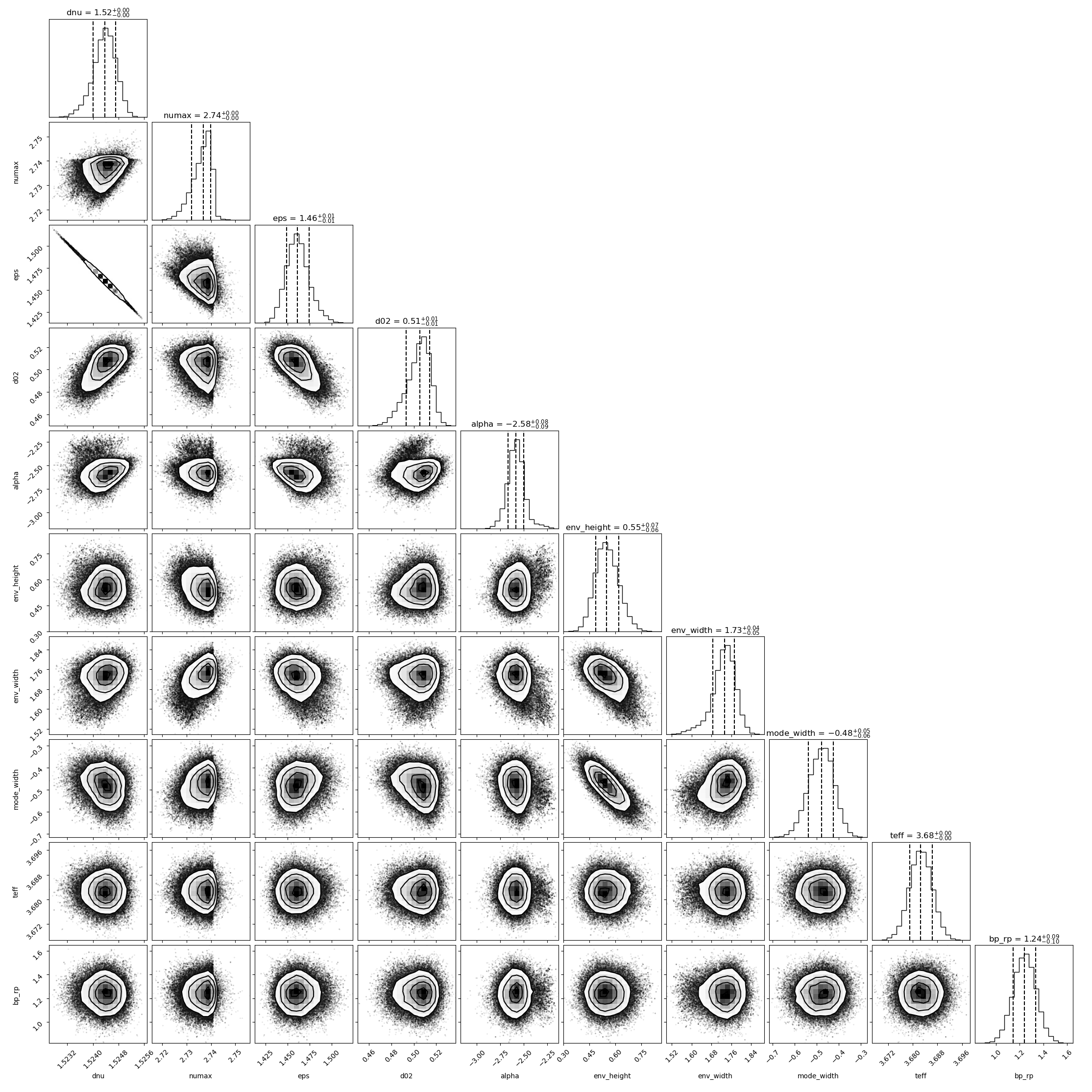}{0.7\textwidth}{Corner diagram of asymptotic fit parameters}
    }
    \caption{Kepler-643 \label{fig:pysyd_corner_643}}
   \end{figure*}

   \renewcommand\thefigure{I}
   \begin{figure*}[ht!]
      \includegraphics[width=0.95\linewidth]{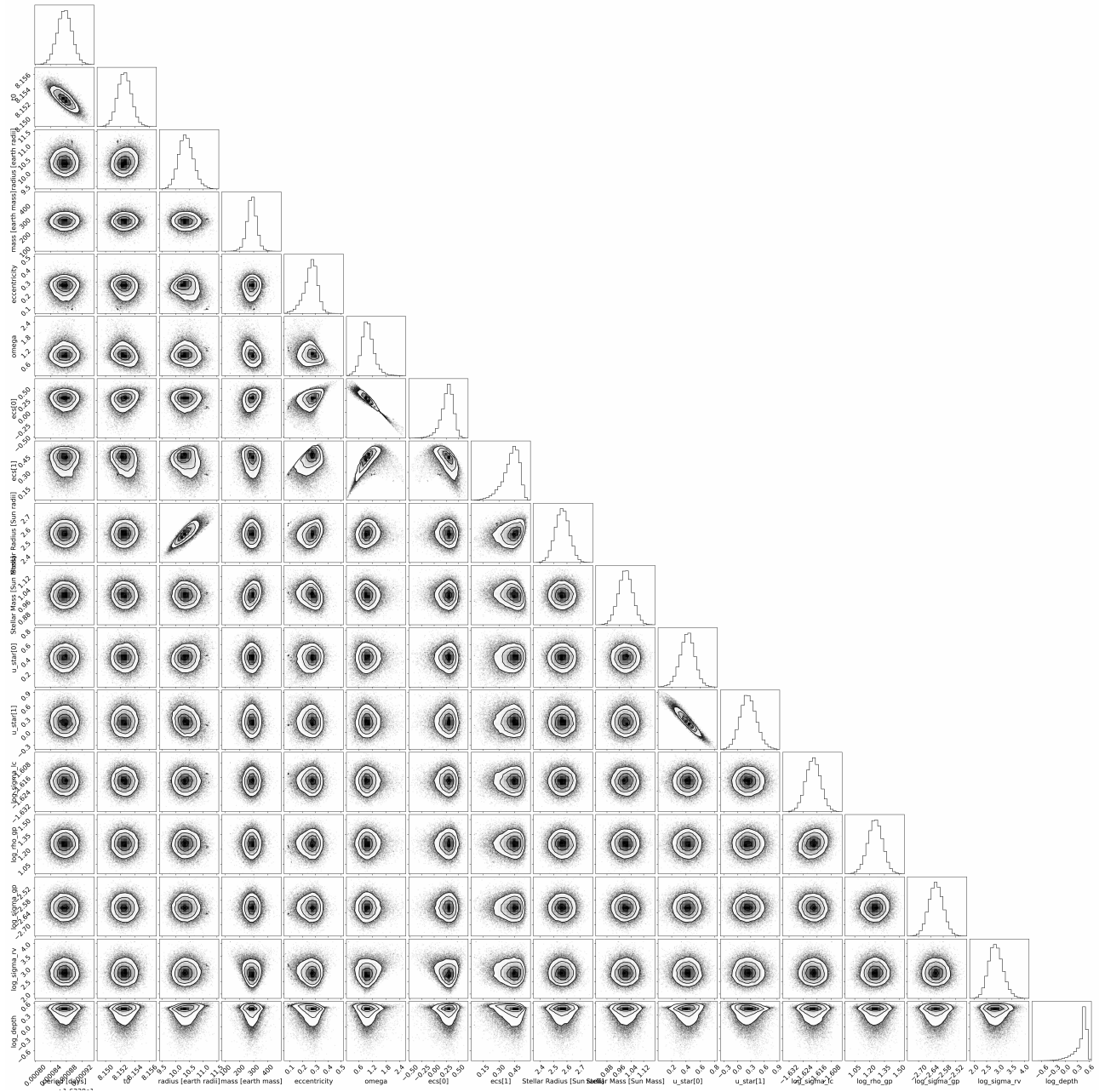}
      \centering
      \caption{Corner diagram of the parameter fitting for Kepler-643 b \label{fig:corner643b}}
   \end{figure*}

\end{document}